\documentclass[traditabstract]{aa}

\usepackage{natbib}
\usepackage{amsmath}
\usepackage{graphicx}
\usepackage{hyperref}  %for hyperlinks
\hypersetup{colorlinks=true, citecolor=blue, linkcolor=blue}

\newcommand{\figpath}{Figures_pdf/}

\newcommand{\tpb}[1]{\textbf{#1}}  

\newcommand{\mi}{J}
\newcommand{\ei}{e}

\begin{document}


\title{Lidov-Kozai stability regions in the $\alpha$\,Cen system}
\titlerunning{Lidov-Kozai stability regions in the $\alpha$\,Cen system}

\author{
C. A. Giuppone\inst{1,2}
\and A. C. M. Correia\inst{2,3}
}

\institute{
Universidad Nacional de C\'ordoba, Observatorio Astron\'omico, IATE, Laprida 854, 5000 C\'ordoba, Argentina 
  \and 
CIDMA, Departamento de F\'isica, Universidade de Aveiro, Campus de
Santiago, 3810-193 Aveiro, Portugal
\and 
ASD, IMCCE-CNRS UMR8028, Observatoire de Paris, 77 Av. Denfert-Rochereau, 75014 Paris, France  
%IMCCE, Observatoire de Paris - PSL Research University, UPMC Univ. Paris 06, Univ. Lille 1, CNRS, 77 Avenue Denfert-Rochereau, 75014 Paris, France
}

\date{\today}

\abstract{
The stability of planets in the $\alpha$\,Centauri\,AB stellar system has been studied extensively.
However, most studies either focus on the orbital plane of the binary or consider inclined circular orbits.
Here, we numerically investigate the stability of a possible planet in the $\alpha$\,Centauri\,AB binary system for S-type orbits in an arbitrary spatial configuration. 
In particular, we focus on inclined orbits and explore the stability for different eccentricities and orientation angles.
We show that large stable and regular regions are present for very eccentric and inclined orbits, corresponding to libration in the Lidov-Kozai resonance.
 {We additionally show that these extreme orbits can survive over the age of the system, despite the effect of tides.}  
Our results remain qualitatively the same for any compact binary system.   
}



\maketitle
    
%________________________________________________________________

\section{Introduction}
The nearest neighbour to our solar system, $\alpha$ Centauri stellar system, always captured the attention of diverse studies in astronomy. The compact binary system $\alpha$\,Cen\,A and $\alpha$\,Cen\,B has an orbital period of 79 yr and has a very eccentric orbit, thus challenging the formation theories of potential existing planets. 
{Planetary accretion models suggest that S-type planets could have formed within the $\alpha$\,Cen system \citep{Quintana_2002,Quintana_2007} provided that the collision velocities of late-stage planetesimals are not too large \citep{Thebault_2008,Thebault_2009,Thebault_2014}. However, the accretional collisions that form planets in compact binary systems is a complicate mechanism, which depends on the initial conditions of the particles and on the mass and orbital parameters of the secondary star \citep[e.g.][]{Beauge_etal_2010}.}  

The announcement of an Earth-mass planet candidate in a 3.24 day orbit $\alpha$\,Cen\,B\,$b$ \citep{Dumusque_2012} and the tentative detection of a transiting planet on a more distant orbit by \cite{Demory_2015} put the spotlight again on this system. 
Nonetheless, \citet{Hatzes_2013} concluded that the presence of the activity signal from the star may boost the velocity amplitude to values comparable to the planet signature. 
Recently, a ground-based radial velocity campaign has ruled out the presence of massive close-in planets \citep{Endl_2015}, although a Jupiter (or less massive) distant planet may exist but has not been detected because the time span of observations is not long enough yet.  

Early works from \cite{Benest_1988, Wiegert_1997} examined the stability of planetary orbits in the $\alpha$\,Centauri system with the logical CPU limitations at the time. 
More recently, \citet{Andrade-Ines_2014, Quarles_2016} have studied the stability regions of this system again.
However, all these studies focused on nearly coplanar orbits or inclined circular orbits. 
Moreover, they have chosen very particular orientation angles for the orbit of the planet, which result in a reduced exploration of the phase space.

In this paper, we study the stability of an additional planet covering all the orbital parameters. 
This allows us to uncover previously unnoticed stability regions.
We describe the methods in Sect.~\ref{sec.methods}. Results are presented in Sects.~\ref{aam}, \ref{sec.results}, and~\ref{stides}, and comparisons with other binary systems are discussed in Sect.~\ref{sec.fictitious}. Finally, our conclusions are presented in Sect.~\ref{sec.conclusions}

\section{Methods}\label{sec.methods}
The orbit of the $\alpha$\,Cen\,AB binary is fully constrained by astrometry. 
Owing to the previous observational constraints, we assumed that if the system is hosting a planet, its mass should be lower than the mass of Jupiter \citep{Endl_2015}. 
Since the two binary stars have similar masses, we studied the motion around the more massive star ($\alpha$\,Cen\,A) using the orbital plane of the binary as a reference frame. 

  \begin{table}
      \caption{Semimajor axis, eccentricity, and masses for the binary systems studied in this paper.}
      \label{tab-bin}
\begin{tabular}{l|c|c|c }
        \hline
& $\alpha$\,Cen\,AB$^{(1)}$  & HD\,196885$^{(2)}$ & HD\,41004$^{(3)}$ \\
        \hline
%$i_B$($^\circ$) & 79.20 $\pm$ 0.041 \\       &    &    
%$\omega_B$($^\circ$) & 231.65 $\pm$ 0.076 \\ &    &    
%$\omega_B$($^\circ$) & 204.85 $\pm$ 0.084 \\ &    &    
$a_B$ (au)      & 23.52            & 21      &  23  \\
$e_B$        & 0.5179 & 0.42      & 0.40   \\
M$_{\rm A}$ (M$_\odot$) & 1.105   & 1.30      & 0.70   \\
M$_{\rm B}$ (M$_\odot$) & 0.934   & 0.45      & 0.40   \\
        \hline
      \end{tabular}
$^{(1)}$ \citet{Quarles_2016}; 
$^{(2)}$ \citet{chauvin_etal_2011}; 
$^{(3)}$ \citet{Zucker_2004, chauvin_etal_2011}.
  \end{table}

Each planetary orbit can be described by six orbital elements: three ``actions'', i.e. the semimajor axis $a$, eccentricity $e$, and inclination $\mi$, with respect to the binary orbital plane; and three ``conjugated'' angles, the mean longitude $\lambda$, longitude of pericentre $\varpi$, and longitude of the ascending node $\Omega$, respectively, measured from the direction of the pericentre of the binary.

Previous works have always focused on the action variables ($a,e,\mi$) and these works arbitrarily fix the conjugated angles, usually at zero ($\lambda=0,\varpi=0,\Omega=0$).
This strategy is understandable. Since the amplitude of the interactions during close encounters depends on action variables, the main features are captured this way.
However, for some particular choices of the angles it is also possible to avoid close encounters, namely when they are involved in resonances (mean motion or secular).
A full exploration of the phase space thus requires the inspection of its six free parameters.
In particular, we need to explore the pairs of conjugated elements, i.e. ($a,\lambda$) to identify regions with mean-motion resonances; and ($e,\varpi$) or ($\mi,\Omega$) to study secular resonances.

We solved the three-body equations of motion numerically with a Burlisch-Stoer integrator with double precision and tolerance $10^{-12}$. We stopped the integrations when the distance of the planet to one of the stars is lower than one stellar radius or when the planet is ejected from the system, identifying the time for any of these situations as ``disruption times''. The integration time is $2 \times 10^5$~yr, which corresponds to several periods of the secular variations.

We analysed the stability of a test planet in a S-type orbit ($a < a_B$) for a wide variety of configurations. 
We constructed stability maps integrating the system on a regular 2D mesh of initial conditions for a pair of orbital parameters, while the remaining four parameters are initially set at nominal values.
Unstable orbits during the integration time, which either collide with one of the stars or escape from the system, are identified in the figures in white.  

For each initial condition we computed the \tpb{Mean Exponential Growth of Nearby Orbits} (MEGNO) value $\langle Y \rangle$ because this value can identify chaotic orbits in less CPU time than other indicators \citep{Cincotta_2000, Maffione_etal_2011}.
We compared this value with other chaos indicators (Lyapunov exponent and frequency analysis) and the chaotic regions coincide.
However, MEGNO cannot give a precise representation of the structure of a resonance, as it only differentiates regular ($\langle Y \rangle \sim 2$, blue regions) from chaotic orbits ($\langle Y \rangle \gg 2$, red regions).

\tpb{To study the structure of the secular resonances, we used the amplitude of maximum variation of the eccentricity of the planet attained during the integrations},  
\begin{equation}
\Delta e= \frac{e_{max}-e_{min}}{2} 
\label{De:chaos} \ .
\end{equation} 
The $\Delta e$ indicator is an extremely useful tool to map the resonant structure in N-body problems \citep{Ramos_2015}; however  $\Delta e$ is not a measure of chaotic motion. Abrupt changes in $\Delta e$ are often traces for the presence of resonances, while regions with large variations in $\Delta e$ are more sensitive to perturbations, thus are very likely chaotic \citep[e.g.][]{Giuppone_etal_2012, marti_giuppone_beauge_2013}. 

\begin{figure}
 \centering
\includegraphics[width=0.9\columnwidth]{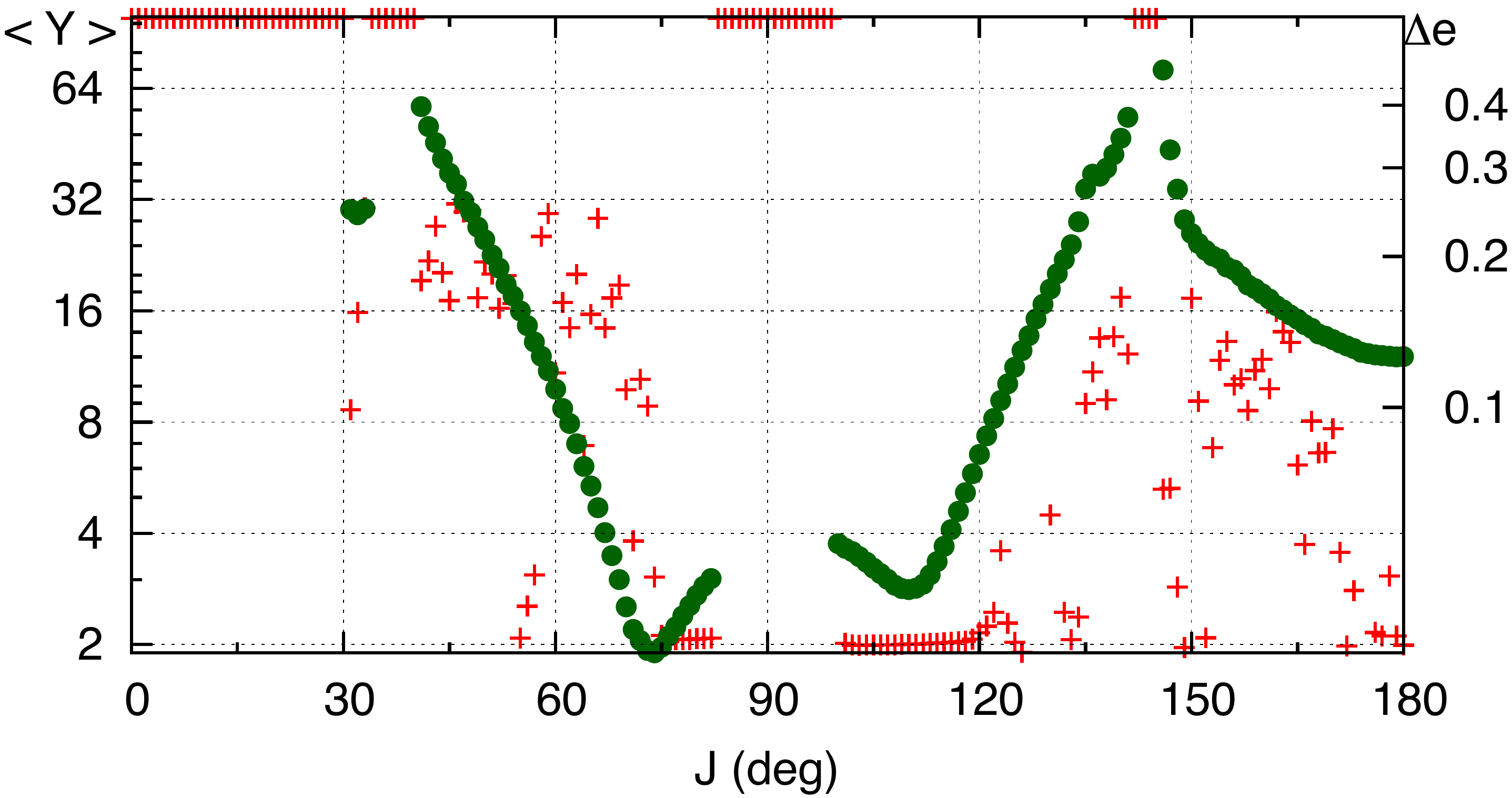} \\
 \caption{Stability indicators for different initial mutual inclination with $a=2$~au, $e=0.9$, $\omega=90^\circ$, and $\lambda=\Omega=0$. 
 The left scale corresponds to the MEGNO chaos indicator, $\langle Y \rangle$ (red crosses), while the right scale corresponds to the $\Delta e$ indicator (green circles). }
 \label{Fig:De}
\end{figure}

\begin{figure}
 \centering
\includegraphics[width=1.\columnwidth]{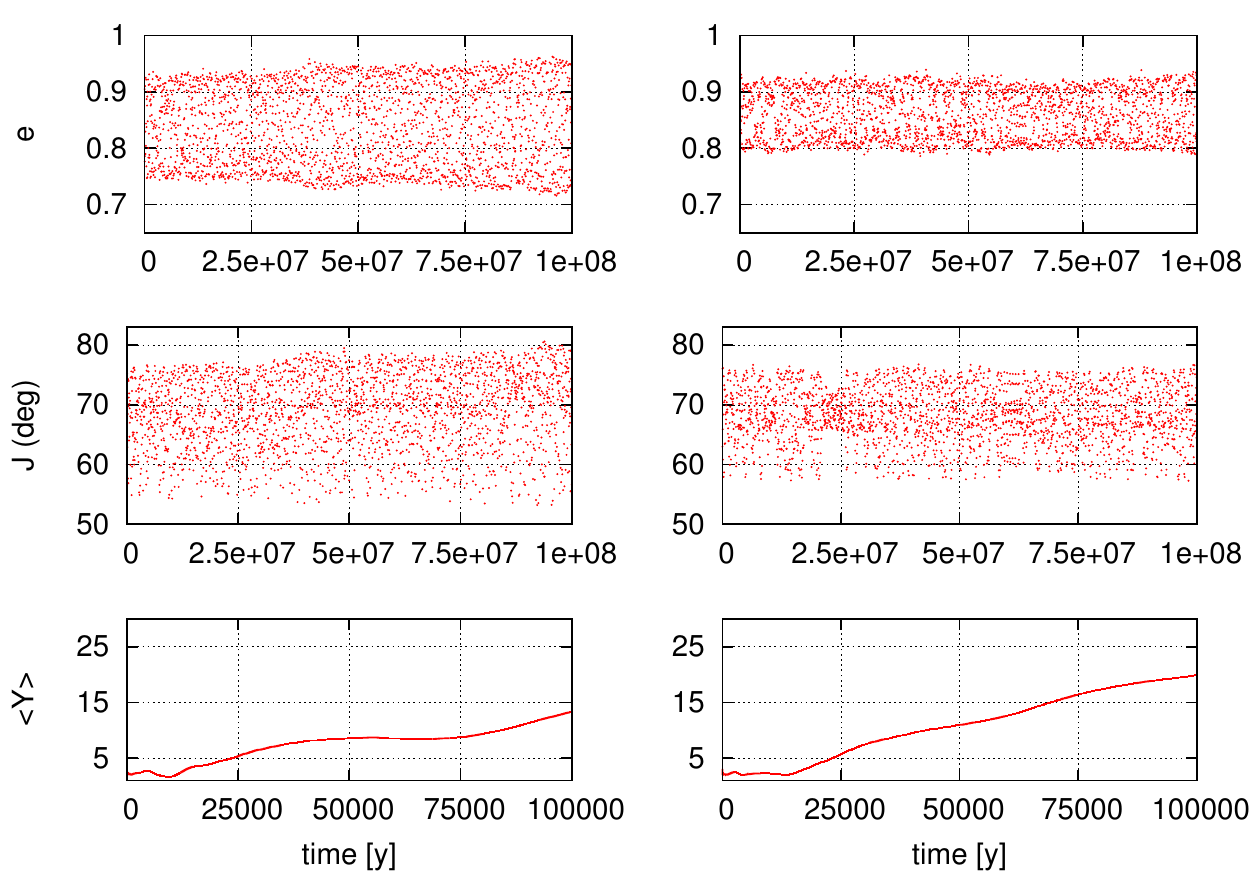} \\
 \caption{Evolution of initial conditions from Fig.~\ref{Fig:De} at $J=60^\circ$ (left) and $J=63^\circ$ (right). Slow diffusion is present in the orbital elements. MEGNO values are proportional to such diffusion (bottom), while $\Delta e$ measures the maximal amplitude eccentricity variations (top). Both orbits survived for $10^9$~yr.}
 \label{Fig:Meg}
\end{figure}

In Figure~\ref{Fig:De} we compare the two indicators used in this work for different values of the initial mutual inclinations. 
{We plot results for $\langle Y \rangle$ and $\Delta e$ values integrated over $10^5$~yr. 
For visual representation, the MEGNO unstable orbits are identified at the top of the scale of $\langle Y \rangle$. 
We can see that the minimum variation of $\Delta e$ corresponds to regions where the orbits are regular $\langle Y \rangle \sim 2$. There are some regions with chaotic orbits ($ 2 < \langle Y \rangle < 16$), that we checked to remain stable over $10^8$~yr with almost the same value of $\Delta e$.} 
In Fig.~\ref{Fig:Meg} we show the evolution of {two bounded chaotic orbits with} the initial conditions at $J=60^\circ$ and $J=63^\circ$.
They undergo slow diffusion in the orbital elements {and still survived over $10^9$~yr}. {Thus, the MEGNO values larger than 2 are somehow related to this diffusion and not necessarily to unstability, i.e. ``stable chaos'' as defined by \citet{Milani_1992}}. 
{Large} MEGNO values give us an estimation of the long-term stability, while $\Delta e$ measures the amplitude of the orbital secular variations.

\section{Action-angle maps}
\label{aam}

\subsection{Mean-motion resonances}

We begin our quest for stability regions by varying the pair ($a, \lambda$), which allow us to identify the presence of possible mean-motion resonances.
We consider coplanar prograde ($\mi=0^\circ$) and retrograde orbits ($\mi=180^\circ$).
We additionally set $\varpi = 0$ and $\Omega=0$, which correspond to orbits with aligned pericentres, as these orbits are among the most favourable for mean-motion resonances to occur.

\begin{figure*}
\centering
\includegraphics[width=0.9\columnwidth]{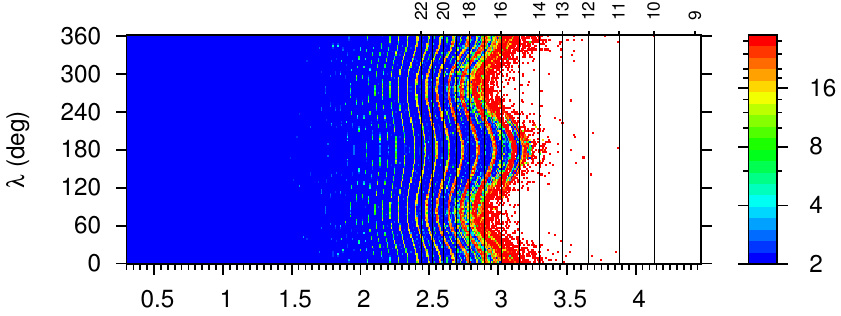} 
\includegraphics[width=0.9\columnwidth]{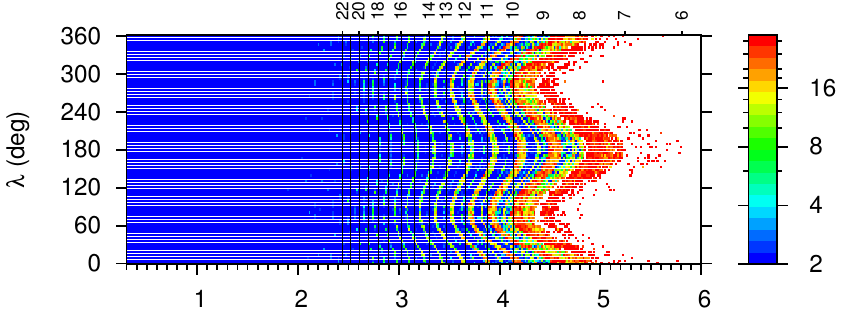} \\
\includegraphics[width=0.9\columnwidth]{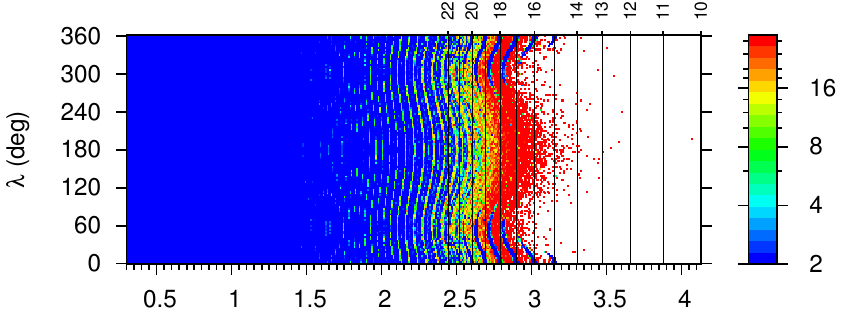} 
\includegraphics[width=0.9\columnwidth]{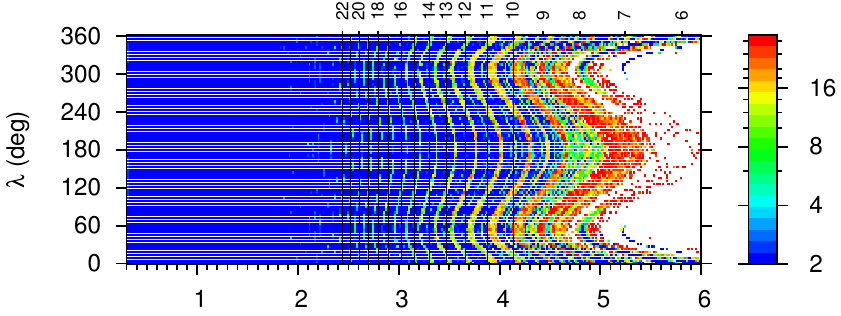} \\
\includegraphics[width=0.9\columnwidth]{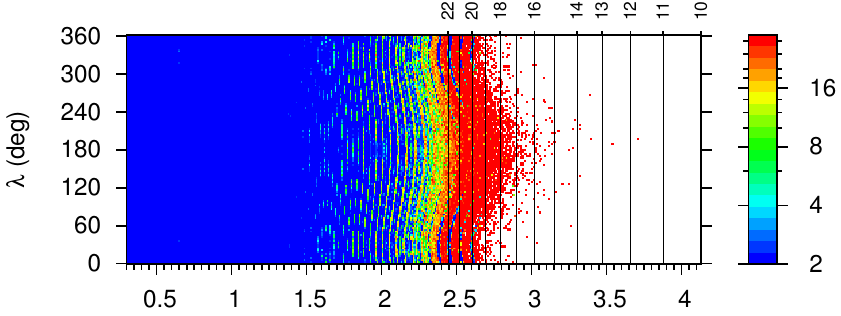} 
\includegraphics[width=0.9\columnwidth]{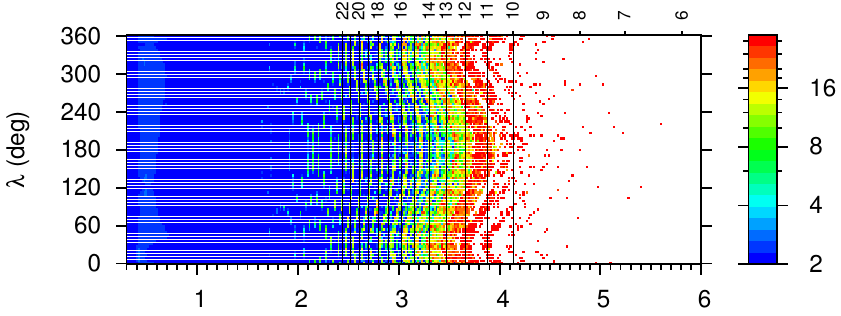} \\
\includegraphics[width=0.9\columnwidth]{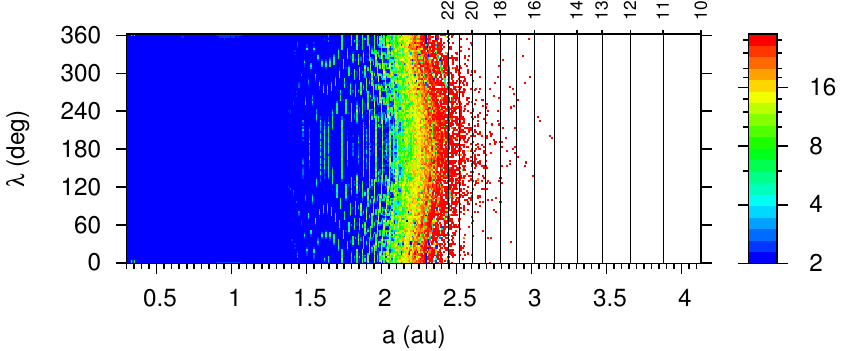} 
\includegraphics[width=0.9\columnwidth]{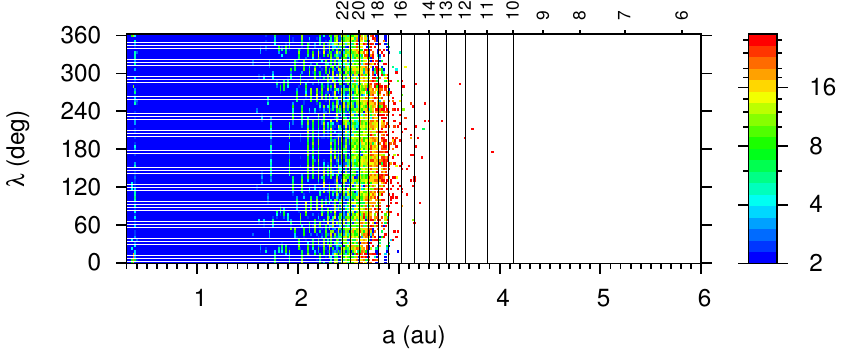} \\
\caption{Stability maps of \textbf{$\langle Y \rangle$} in the plane ($a,\lambda$) for some eccentricity values with $\Omega=\omega=0^\circ$,  $\mi=0^\circ$ (left) and $\mi=180^\circ$ (right). From top to bottom the initial $e=0, 0.3, 0.6,$ and $0.9$. Vertical lines indicate the main mean-motion resonances. 
}
 \label{a:lambda}
\end{figure*}

In Figure~\ref{a:lambda} we show the stability maps for different values of the initial eccentricity, $\ei = 0, 0.3, 0.6$ and $0.9$.
We extend the semimajor axis up to $a=6$~au, since the Hill radius of $\alpha$\,Cen\,A is $\sim 6.4$~au \citep[e.g.][]{Marchal_Bozis_1982}.
{The initial conditions with MEGNO values $\langle Y \rangle \sim 2$ are regular, orbits with $2 < \langle Y \rangle < 16$ show very small diffusion in orbital elements, and orbits with $\langle Y \rangle \gtrsim 16$, identified as red regions in the dynamical maps, show high diffusion and eventually collide with one of the stars.} 

As in all previous studies, we observe that in the prograde case, stable orbits are only possible for $a \lesssim 3$\,au \citep{holman1999AJ}.
Stability for $3 \lesssim a \lesssim 6$~au is not possible because of resonance overlap, which leads to chaotic motions in these regions \citep{wisdom1980aj}.
Although stability slightly depends on the initial $\lambda$ value, we see that it is not possible to trap a planet in a low order mean-motion resonance with the companion star.
However, for small eccentricities we can observe some small resonant islands for the 15:1 and the 16:1 mean-motion resonances.

{\cite{Marzari_2016} have used frequency map analysis to show the stability of a planet in a binary system with $e_B=0$ and $e_B=0.4$. 
These authors have shown that in the case of an eccentric binary, for a given semimajor axis the orbit of the planet can be regular or chaotic, depending on the initial mean longitude, $\lambda$. 
However, their initial $\lambda$ values were chosen randomly, so it was not possible to understand the origin of this behavior exactly.
In Fig.~\ref{a:lambda} we clearly see that for $\lambda$ close to $0^\circ$ and $180^\circ$ stable resonant islands exist at the middle of chaotic regions, which allow different stability regimes for the same semimajor axis value.}

For retrograde orbits we observe that stability is possible for larger values of the semimajor axis.
In particular, for $\ei=0.3$ stability is possible up to 6~au, very close to the Hill sphere.
We additionally observe that capture in lower order mean-motion resonances, such as  7:1 or 6:1, is also possible.
Indeed, retrograde orbits in binary systems are more stable than the prograde orbits because of a different structure of mean-motion resonance overlaps \citep[see][]{Morais_2012}.
We hence conclude that the inclination value is a very important parameter that shapes the stability in binary systems.

\subsection{Secular resonances}

\begin{figure*}
 \centering     
\includegraphics[height=4.6cm]{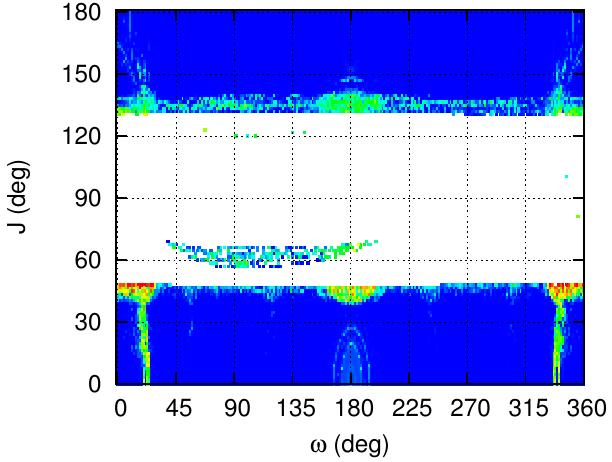}  
\includegraphics[height=4.6cm]{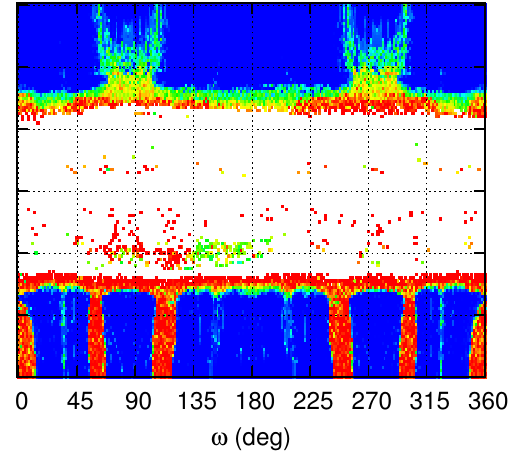} 
\includegraphics[height=4.6cm]{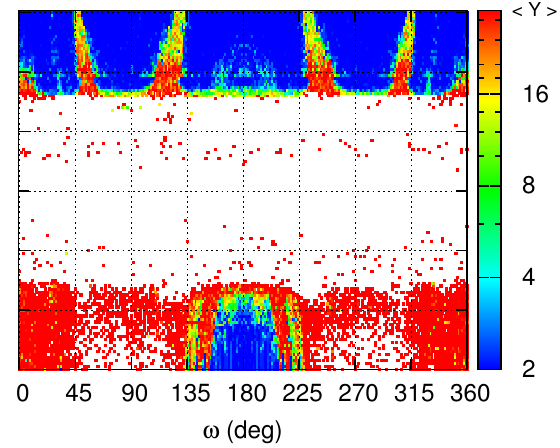}\\ 
\includegraphics[height=4.6cm]{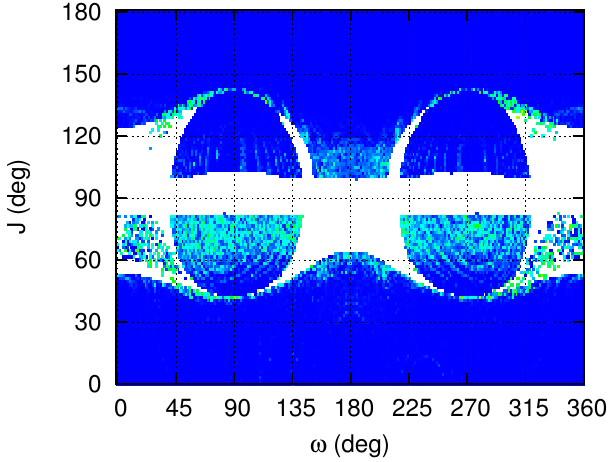}  
\includegraphics[height=4.6cm]{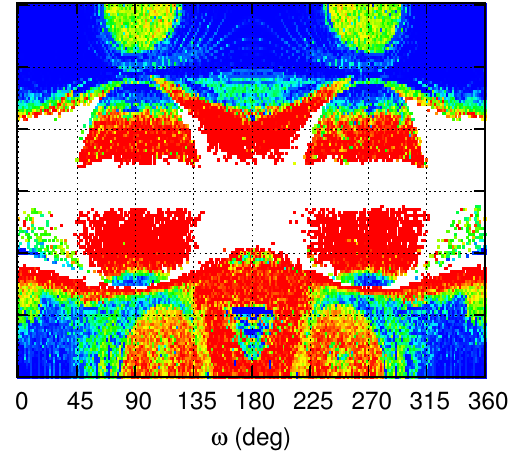} 
\includegraphics[height=4.6cm]{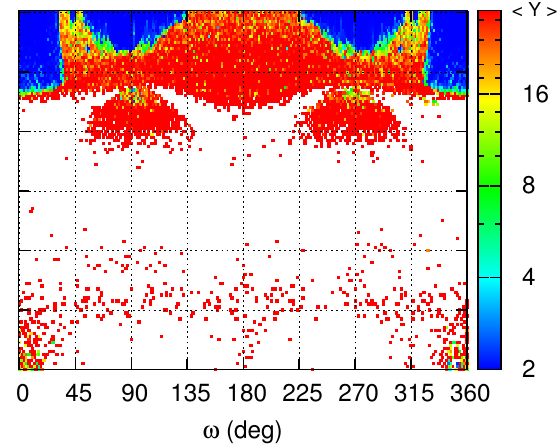}\\
\includegraphics[height=4.6cm]{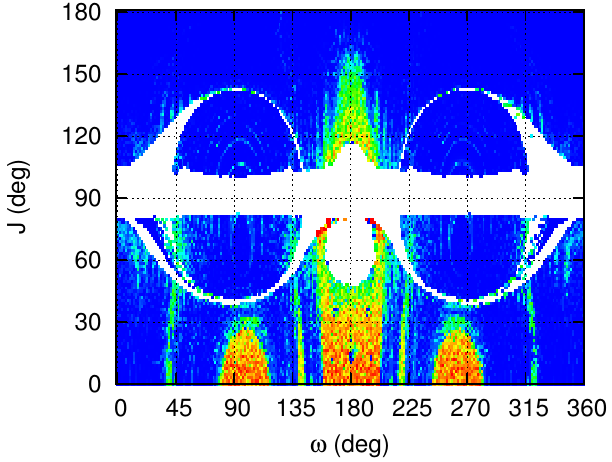}  
\includegraphics[height=4.6cm]{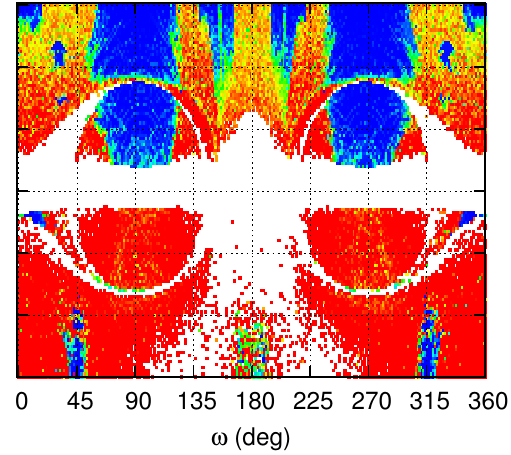} 
\includegraphics[height=4.6cm]{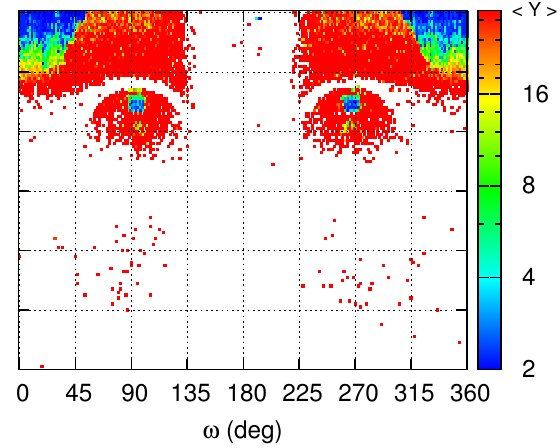}\\
\includegraphics[height=4.6cm]{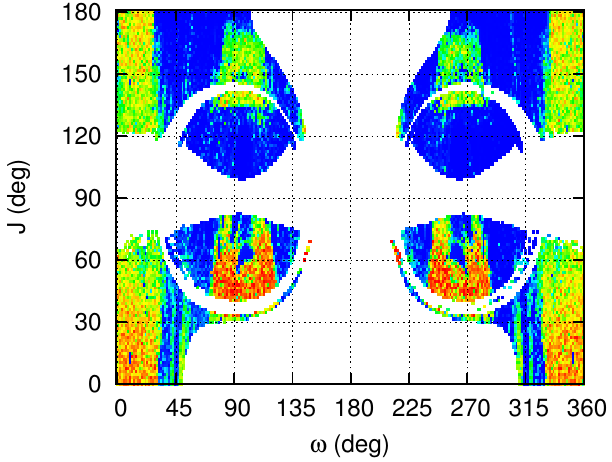}  
\includegraphics[height=4.6cm]{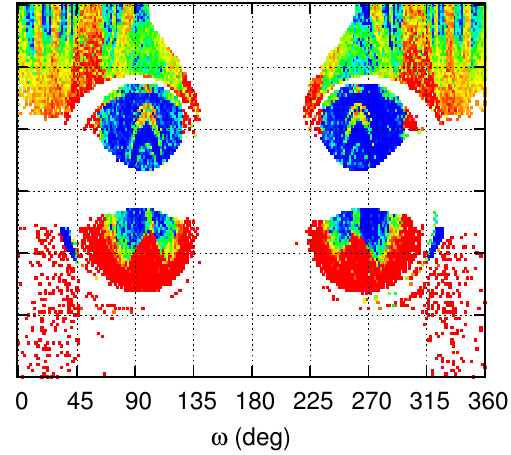} 
\includegraphics[height=4.6cm]{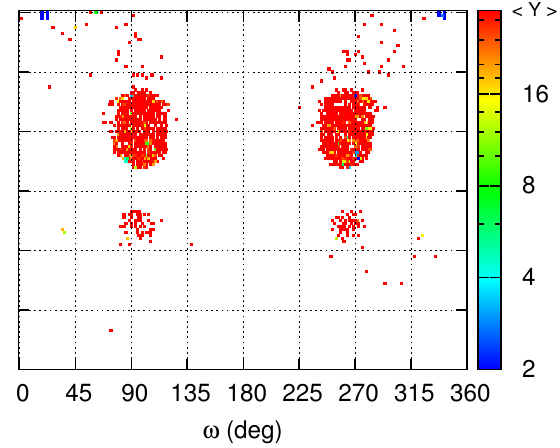}\\
 \caption{Stability maps of \textbf{$\langle Y \rangle$} in the plane ($\omega, \mi$) for some semimajor axis and eccentricity values with $\lambda=\Omega=0$. From left to right the initial $a=2$~au, $a=2.42$~au, and $a=3$~au. From top to bottom the initial $e=0$, $0.3$, $0.6$, and $0.9$.}
 \label{fig-MapI}
\end{figure*}

The main frequency involved in the angle $\lambda$ is the orbital mean motion, $n$, thus this angle varies rapidly.
The angles $\Omega$ and $\varpi$ vary in a much longer timescale owing to the presence of the binary companion, hence the name secular.
Since we are studying a three-body problem, owing to the conservation of the total angular momentum there is a single frequency associated with the precession of the line of the nodes, $s \approx \dot \Omega$.
Moreover, since most of the angular momentum is on the binary orbit, the precession frequency associated with the pericentre of the binary is almost zero. 

Thus, the longitude of the pericentre is mainly driven by a single frequency $g \approx \dot \varpi$, which corresponds to the precession rate of the pericentre of the planet.
Indeed, in the restricted problem, $g$ and $s$ are the only secular frequencies in the system.
As a consequence, secular resonances can only occur when $g = s$, which is usually known in the literature by the Lidov-Kozai resonance \citep{Lidov_1962, Kozai_1962}.
The particular geometry of this almost restricted three-body problem allows us to explore the phase-space more rapidly. 
Instead of using the angles $\Omega$ and $\varpi$ separately, we can adopt the argument of the pericentre $\omega = \varpi - \Omega$, for which resonances occur when $\dot \omega = g - s = 0$.
Therefore, all the significant information on secular resonances can be captured by a $(\omega, e)$ or $(\omega, \mi)$ diagram.

In Figure~\ref{fig-MapI} we show the stability maps in the plane $(\omega, \mi)$ for different values of the initial eccentricity ($\ei = 0$, $0.3$, $0.6$ and $0.9$), and {three different values of the semimajor axis, corresponding to three different stability regions}:
$a=2$~au, which places the planet inside a stable region for prograde orbits; $a=2.42$~au, corresponding to the 22:1 mean-motion resonance and near the unstable region; and $a=3$~au, already in a chaotic region for prograde orbits (see Fig\,\ref{a:lambda}).

{In Figure~\ref{fig-MapI}} we also observe that polar orbits ($\mi \sim 90^\circ$) are always unstable. 
Owing to the conservation of the orbital angular momentum, the following quantity is conserved:
\begin{equation}
%h \equiv 
\sqrt{(1-e^2)} \cos \mi = cte \ .
\label{h:cte}
\end{equation}
As a consequence, the eccentricity of polar orbits can reach values very close to unity, which may place the planet outside the Hill sphere or in collision with the star.

For initial circular orbits ($\ei=0$) stability is only possible for nearly coplanar orbits ($\mi \lesssim 30^\circ$ or $\mi \gtrsim 150^\circ$). 
Prograde orbits ($\mi \lesssim 30^\circ$) become unstable for $a \sim 3$~au, in conformity with the results shown in Fig.~\ref{a:lambda}, except for a small zone of aligned and anti-aligned orbits ($\omega \approx 0^\circ$ and $\omega \approx 180^\circ$, respectively).
However, as also shown in Fig.~\ref{a:lambda}, the stability region is extended beyond this value of the semimajor axis for retrograde orbits ($\mi \gtrsim 150^\circ$).
{Some additional chaotic structures can also be seen for some $\omega$ values, probably due to secondary non-linear resonances.}

It is often assumed that circular orbits provide an upper limit for stability with a given semimajor axis, since the minimal distance between the planet and the perturber decreases with the eccentricity.
However, we observe that as we increase the initial eccentricity, the coplanar regions become indeed less stable, but new stable islands emerge in the region within $30^\circ \lesssim \mi \lesssim 150^\circ$.
An interesting result is that for $a>2$~au stable prograde coplanar orbits are no more possible for moderate eccentricities, but stability can still be achieved in these islands for very high values of eccentricity and mutual inclination (Fig.~\ref{fig-MapI}).

The stability islands at high inclinations are centred at $\omega = 90^\circ$ and $\omega = 270^\circ$ and correspond to the secular Lidov-Kozai resonances.
For the restricted problem, the resonant motion is possible whenever  \citep{Lidov_1962, Kozai_1962}
\begin{equation}
3 (1-e^2) \ge 5 \cos^2 \mi 
 \ . \label{h:lkr0}
\end{equation}
We then conclude that resonant motion is only possible for $\mi_c \le \mi \le \pi - \mi_c $, with the critical inclination $\mi_c = \cos^{-1} \sqrt{3/5}  \approx 39.2^\circ$  (corresponding to zero eccentricity).
The equality above also corresponds to exact resonance.
For a given initial eccentricity $\ei$, the equilibrium inclination $\mi_r$ at exact resonance is then given by
\begin{equation}
 \cos \mi_r = \sqrt{3/5} \sqrt{1-\ei^2}
 \ . \label{h:lkr} 
\end{equation}

For instance, for $\ei=0.6$ we have $\mi_r = 51.7^\circ$ or $\mi_r = 128.3^\circ$ and for $\ei=0.9$ we have $\mi_r = 70.3^\circ$ or $\mi_r = 109.7^\circ$.
For trajectories in libration around this equilibrium, the minimum value of the inclination for prograde orbits lies in the interval $ \mi_c \le \mi < \mi_r$ and the maximum inclination in retrograde orbits lies in $ \pi - \mi_c \ge \mi > \pi - \mi_r$.
Since for initial circular orbits ($\ei=0$) we have $\mi_c = \mi_r$, the entire region occupied by the Lidov-Kozai resonance is unstable.
However, as we increase the initial eccentricity, the equilibrium mutual inclination for prograde (retrograde) orbits moves to higher (lower) values and stable resonant regions appear in the vicinity of $\omega = 90^\circ$ and $\omega = 270^\circ$.

In Figure~\ref{Fig:De} we show the value of the chaotic indicators for a vertical line with $\omega=90^\circ$ in the bottom left panel of Fig.~\ref{fig-MapI} ($a=2$~au and $e=0.9$).
We observe that the most {regular region} is obtained for the libration regions of the Lidov-Kozai resonance. 
{Regular} coplanar retrograde orbits are still possible in this case, but they already present eccentricities very close to instability.

\section{New action-action maps}
\label{sec.results}

Maps involving the three actions ($a,e,\mi$) were extensively explored in previous studies \citep[e.g.][]{Andrade-Ines_2014, Quarles_2016}.
However, they usually fix $\omega = 0^\circ$, which corresponds to a region of the phase-space that is always outside the libration zone of the Lidov-Kozai resonance (Fig.~\ref{fig-MapI}).
As we just saw in previous section, stable regions for high eccentricity and mutual inclination are near the centre of libration, which is placed at $\omega = 90^\circ$ or $\omega = 270^\circ$.
Therefore, it is better to fix $\omega$ at one of these two values to capture the resonant regions in action-action maps.

\begin{figure*}
\centering
\includegraphics[width=0.75\columnwidth]{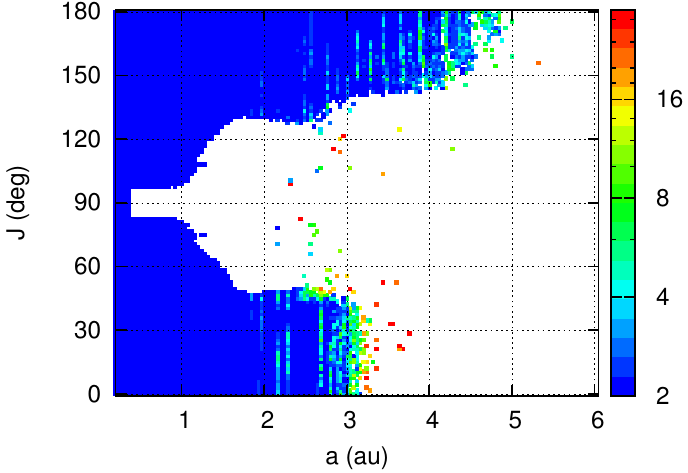}
\includegraphics[width=0.75\columnwidth]{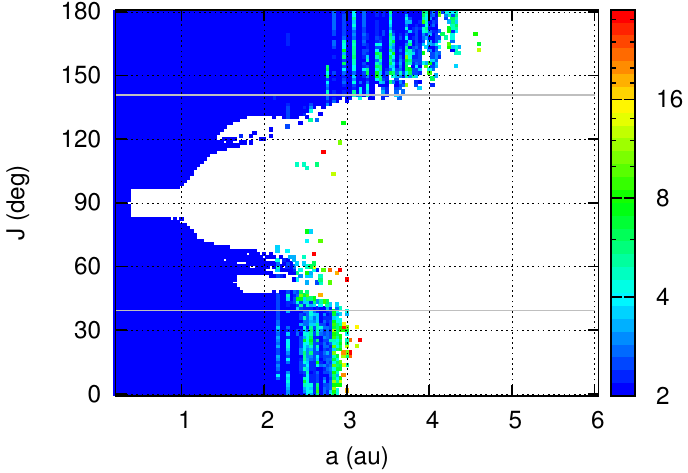} \\
\includegraphics[width=0.75\columnwidth]{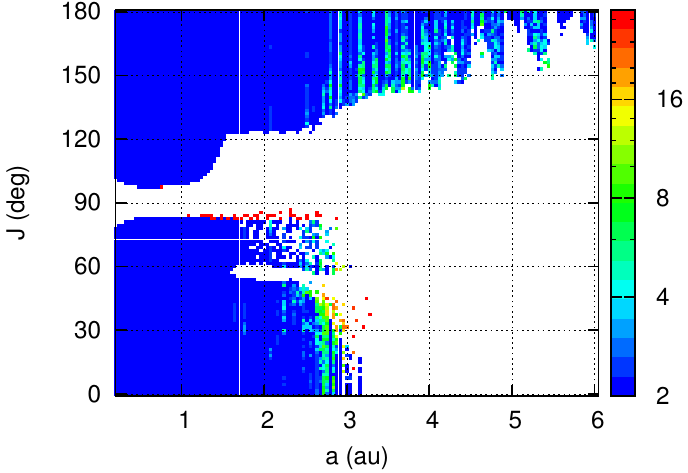}
\includegraphics[width=0.75\columnwidth]{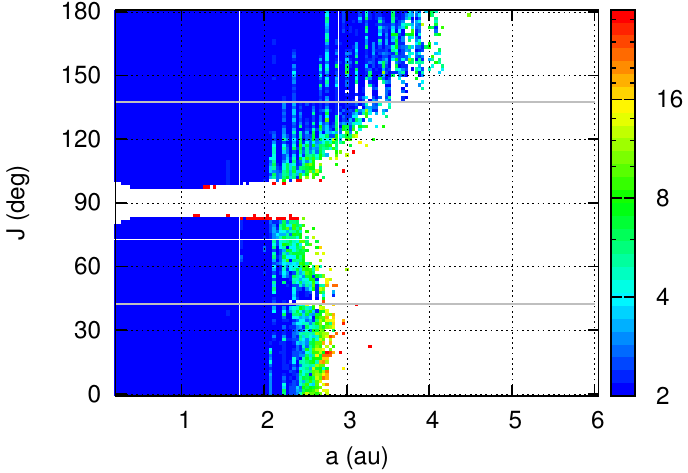} \\
\includegraphics[width=0.75\columnwidth]{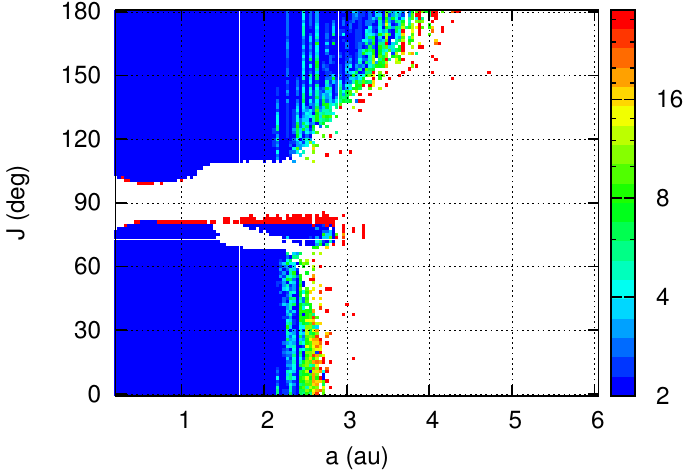} 
\includegraphics[width=0.75\columnwidth]{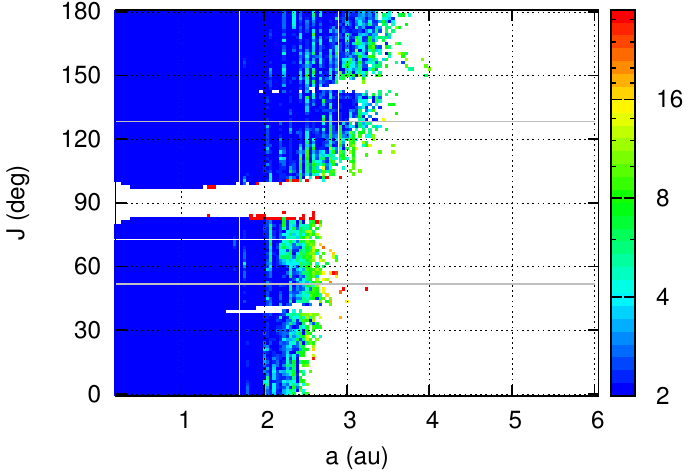} \\
\includegraphics[width=0.75\columnwidth]{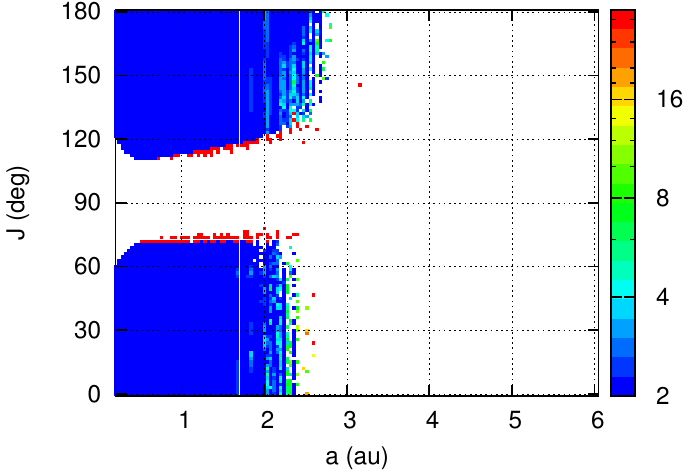} 
\includegraphics[width=0.75\columnwidth]{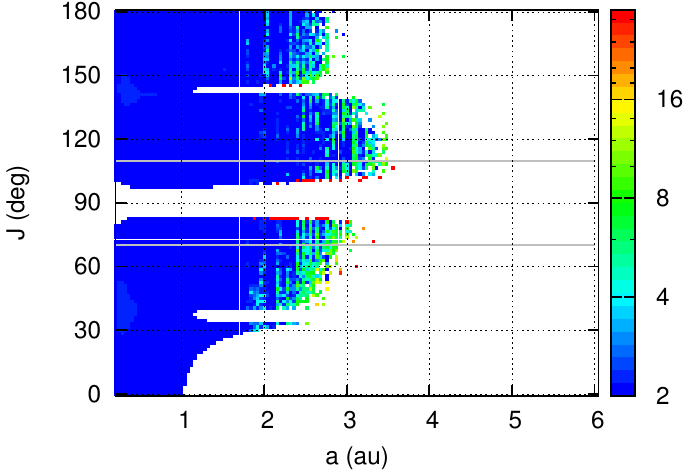} \\
\caption{Stability maps of \textbf{$\langle Y \rangle$} in the plane ($a, \mi$) for some eccentricity values with $\lambda=\Omega=0^\circ$ and $\omega=0^\circ$ (left) or $\omega=90^\circ$ (right). From top to bottom the initial eccentricity is $e=0$, $0.3$, $0.6$, and $0.9$.
{The horizontal grey lines give the centre of libration of the Lidov-Kozai resonance (Eq.\,(\ref{h:lkr}))}.}
\label{fig-LK}
\end{figure*}

\begin{figure*}
 \centering
\includegraphics[width=\columnwidth]{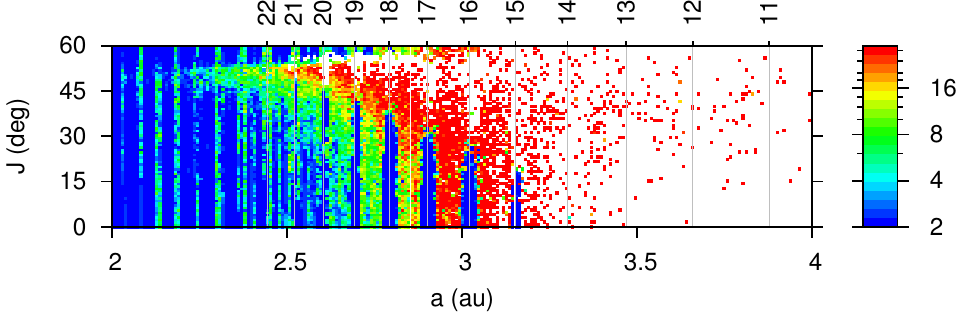}
\includegraphics[width=\columnwidth]{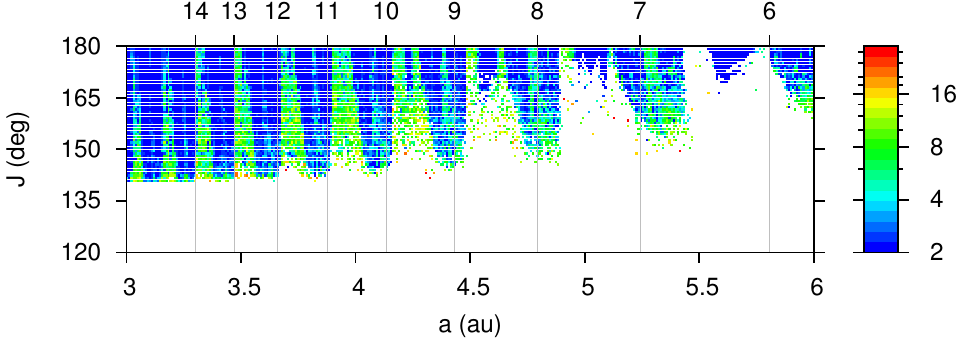}
 \caption{Stability maps of \textbf{$\langle Y \rangle$} in the plane ($a, \mi$) for $\ei=0.3$ and $\omega=0^\circ$ with $\lambda=\Omega=0$. The left panel shows prograde orbits and right panel for retrograde orbits. The vertical labels indicate the nominal position of the $N$:1 mean-motion resonances. }
 \label{fig-LKz}
\end{figure*}

\subsection{Semimajor axis versus mutual inclination}

In Figure~\ref{fig-LK} we show the stability maps in the plane $(a, \mi)$ for different values of the initial eccentricity $\ei = 0$, $0.3$, $0.6,$ and $0.9$.
We fix $\lambda=\Omega=0$, and $\omega=0^\circ$ (left column) or $\omega=90^\circ$ (right column) to compare better with previous studies.
We can see that for both $\omega$ values the region around polar orbits is very chaotic and splits the regions corresponding to prograde ($\mi<90^\circ$) and retrograde ($\mi>90 ^\circ$) orbits. Generally, collisions and/or ejections occur in less than $5 \times 10^4$~yr for nearby orbits. 
We scale our integrations time such that more distant orbits are integrated at least $5 \times 10^4$ planetary periods. 

For initial circular orbits ($\ei=0$) there is no big difference between the two $\omega$ values.
Indeed, in the restricted quadrupolar problem, the Hamiltonian only depends on the product $e^2 \cos 2 \omega$ \citep[e.g.][]{Kozai_1962, Giuppone_etal_2012}, so the initial value of $\omega$ does not change the energy at the quadrupole order,  which is the dominating term.
We also observe that retrograde coplanar orbits are stable for larger values of semimajor axis than prograde coplanar orbits \citep[as also noted by][]{holman1999AJ, Quarles_2016}.
Indeed, for compact binary systems the retrograde planets are stable up to distances closer to the perturber than prograde planets owing to the higher order of overlap of nearby resonances \citep{Morais_2012}.

As we increase the initial eccentricity, the coplanar orbits remain more stable in the case $\omega=0^\circ$, as they correspond to aligned orbits \citep[see][]{giuppone_morais_correia_2013}.
However, for inclined orbits we observe that a new stability region appears for maps with $\omega=90^\circ$, in a strip for mutual inclinations given by expression (\ref{h:lkr}), corresponding to libration in the Lidov-Kozai resonance.
We also observe a chaotic strip clearly delimiting the coplanar and resonant regions corresponding to the separatrix of this resonance.

In Figure~\ref{fig-LK} we see that the eccentricity does not limit the stability in binary systems, provided that we change the mutual inclination following the resonant equilibrium points (Eq.\,(\ref{h:lkr})).
Moderate eccentricities can also facilitate stability for coplanar orbits.
Indeed, for $e=0.3$ and $\omega = 0^\circ$ (aligned orbits) we observe that stable regions exist beyond 3~au and 6~au for prograde and retrograde orbits, respectively.
In Figure~\ref{fig-LKz} we zoom in on the coplanar regions for these initial conditions. 
We superimposed the nominal location of the $N$:1 mean-motion resonances to better understand the structures present in these regions. 

We observe that stability islands are associated with mean-motion resonances between the planet and the stellar companion $\alpha$\,Cen\,B.
The last stable resonances correspond to the 15:1 for prograde orbits and 6:1 for retrograde orbits.
Lower order resonances are not possible because they lie outside the Hill sphere of $\alpha$\,Cen\,A.
The only real limitation for stability is thus the semimajor axis; for the remaining orbital parameters stability can always be achieved at some particular combinations.

\subsection{Eccentricity versus mutual inclination}

\begin{figure*}
 \centering
\includegraphics[height=6.9cm]{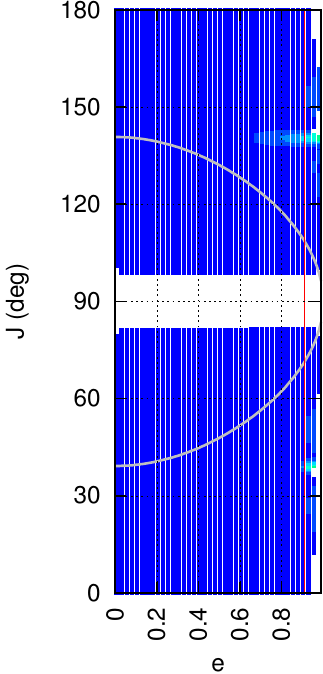} 
\includegraphics[height=6.9cm]{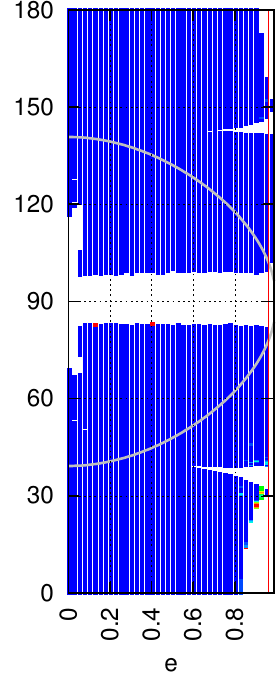}
\includegraphics[height=6.9cm]{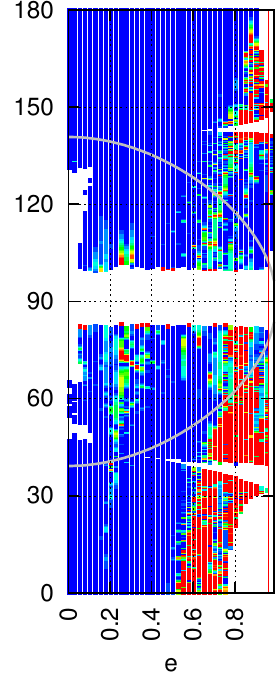} 
\includegraphics[height=6.9cm]{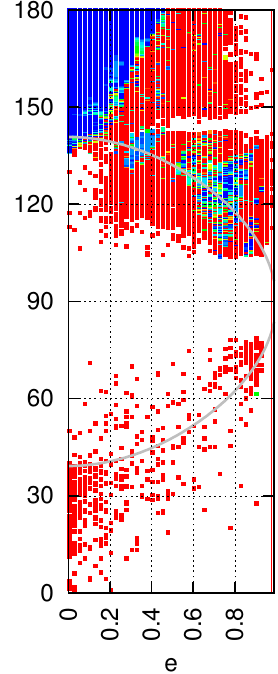} 
\includegraphics[height=6.9cm]{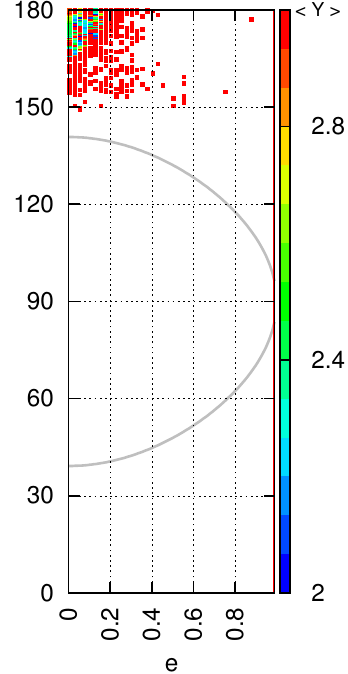}\\
\includegraphics[height=6.9cm]{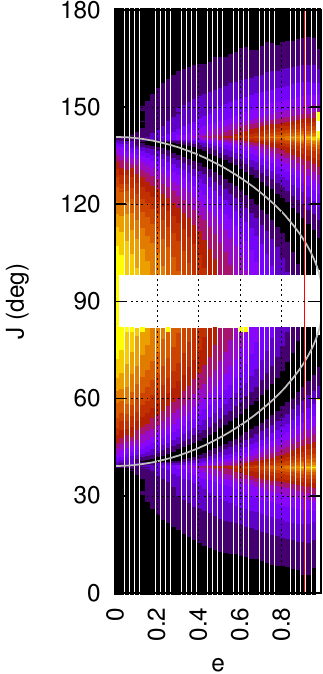} 
\includegraphics[height=6.9cm]{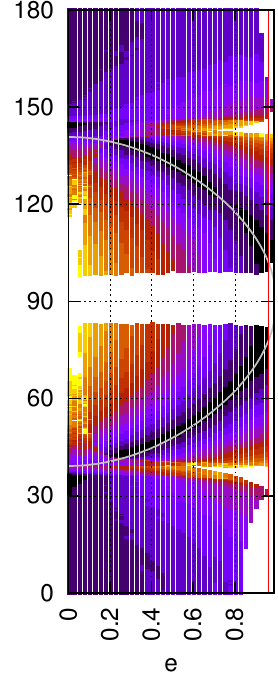}
\includegraphics[height=6.9cm]{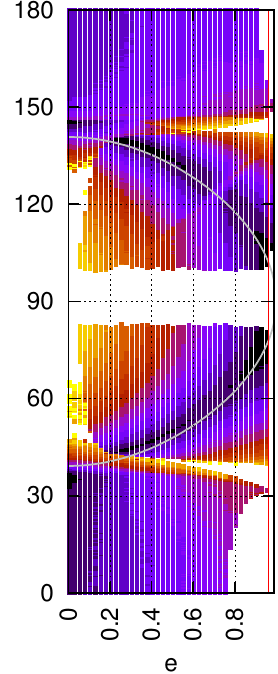} 
\includegraphics[height=6.9cm]{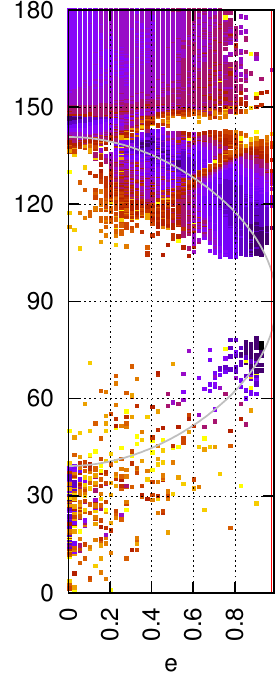} 
\includegraphics[height=6.9cm]{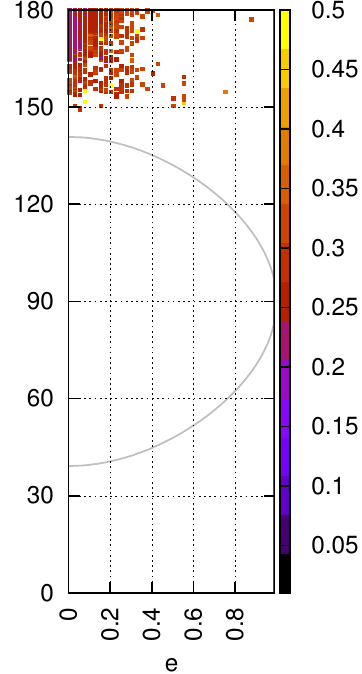}\\
\caption{Stability maps in the plane ($e, \mi$) with $\lambda=\Omega=0$ and $\omega=90^\circ$ for initial semimajor axis from left to right: $a=0.55$, 1.5, 2.0, 3.0, and 4.0~au.
We show the MEGNO $\langle Y \rangle$ (top) and the $\Delta e$ (bottom) stability indicators.
The grey curve gives the centre of libration of the Lidov-Kozai resonance (Eq.\,(\ref{h:lkr})), while the red vertical line gives the limit for tidal stability (Eq.\,(\ref{100210d})).}
 \label{fig-MapEI}
\end{figure*}

In Figure~\ref{fig-MapEI} we show the stability maps in the plane $(e, \mi)$ for different values of the  semimajor axis $a = 0.55$, $1.5$, $2.0$, $3.0,$ and $4.0$~au.
We fix $\lambda=\Omega=0$ and $\omega=90^\circ$ such that the Lidov-Kozai resonance is visible.
We also plot the curve corresponding to the centre of this resonance, given by expression (\ref{h:lkr}) obtained in the frame of the restricted quadrupolar approximation.

We observe that for small values of the semimajor axis ($a < 2$~au), almost all configurations are stable, except for those in coplanar orbits.
However, for the semimajor axis of roughly 3~au, most configurations become unstable.
As in previous figures, we see that only a small retrograde zone subsists for small eccentricities, together with the Lidov-Kozai regions.
In particular, there is a perfect agreement between the theoretical prediction given by expression (\ref{h:lkr}) and the stable regions with high inclination.

In Fig.~\ref{fig-MapEI} (bottom) we also show the $\Delta e$ stability indicator (Eq.\,(\ref{De:chaos})) to get a clearer view of the secular dynamics in the $\alpha$\,Cen system.
We observe that the libration resonant areas are the most stable structures in the system.
\tpb{Libration regions are present for close-in orbits ($a<1$~au) together with the stable coplanar regions, but the former subsist for more distant semimajor axes ($a \sim 3$~au), while the coplanar regions are no longer stable.}
 The stability regions are also larger for high values of eccentricity and inclination, since the libration zone is more extended.

\section{Tidal evolution}
\label{stides}

For some orbital configurations, the eccentricity of the planet may reach very high values and become close enough to the central star at periastron to undergo tidal effects.
In that case, the semimajor axis and the eccentricity will decrease and the final configuration can be completely different from the initial configuration.

For an unperturbed orbit,
the secular evolution of the eccentricity by tidal effect using a linear dissipation model can be given by \citep{Correia_2009}
\begin{equation}
\dot e = - K_0 f (e) \, e \ , \label{100210b} 
\end{equation}
with
\begin{equation}
f (e) = \frac{1 + \frac{45}{14}e^2 + 8e^4 + \frac{685}{224}e^6 + \frac{255}{448}e^8 + \frac{25}{1792}e^{10}}{\left(1 + 3e^2 + \frac{3}{8}e^4\right) (1-e^2)^{-3/2}}  \ , \label{090527a}
\end{equation}
and
\begin{equation}
K_0 = n_0  \frac{21}{2} \frac{k_2}{Q} \frac{M_A}{m} \left(\frac{R}{a_0}\right)^5 (1-e_0^2)^{-8} \ ,
\label{100210c} 
\end{equation}
where $n_0=\sqrt{G M_A /a_0^3}$ is the initial mean motion, $k_2 $ is the second Love number, $Q$ is the tidal dissipation factor, $m$ is the mass of the planet, and $R$ its radius.

The solution of the above equation is given by \citep{Correia_Laskar_2010B}
\begin{equation}
F(e) = F(e_0) \exp(- K_0 t) \ , \label{100210d} 
\end{equation}
where $ F(e) $ is an implicit function of $ e $, which converges to zero as $ t
\rightarrow + \infty $.
The characteristic timescale for fully dampening the eccentricity of the orbit
is then $ \tau \sim 1 / K_0 $.
Orbits with $\tau $ smaller than the age of the system can be excluded from the stability diagrams because the planet will not stay \tpb{at the original semimajor axis 
because the orbit of the planet evolves.}
This does not mean that the planet is necessarily destroyed, only that it evolves into a different value of the semimajor axis and/or eccentricity.

The time $\tau$ depends on many uncertain parameters, so it is not easy to place a clear limit for tidal stability. 
In particular, $\tau$ should be different for rocky and gaseous planets, since rocky bodies usually dissipate energy more efficiently.
Indeed, rocky planets in the solar system present $k_2/Q \sim 10^{-2} - 10^{-3} $, while for gaseous planets $k_2/Q \sim 10^{-4} - 10^{-5} $ \citep{Yoder_1995cnt}.
However, the mass and radius of a gaseous planet is in general $10^2$ and $10$ times larger than mass and the radius of a rocky planet, respectively.
When replacing all these values in expression (\ref{100210c}) we get similar values for $\tau$ for both types of planets.

In Figure~\ref{fig-MapEI} we trace a vertical red line corresponding to the solution of equation (\ref{100210d}) for a timescale $\tau$ smaller than 1~Gyr assuming
%a Saturn-like planet with $k_2/Q = 2.3 \times 10^{-4}$ \citep{Lainey_etal_2012}:
a Jupiter-like planet with $k_2/Q = 1.1 \times 10^{-5}$ \citep{Lainey_etal_2009};
%\begin{equation}
%F(\ei) = F(0.1) \exp (K_\mathrm{Gyr^{-1}}) \ . \label{170103a} 
%\end{equation} 
the solutions with higher initial eccentricities should not be considered, since they evolve in a period of time shorter than the age of the system.
As expected, we observe that for smaller values of semimajor axis we have to exclude more configurations, since tides are stronger and the orbits evolve faster.
However, for $a = 1.5$~au, we only need to exclude orbits with $\ei > 0.96$. 
For the Lidov-Kozai resonance, these eccentricities correspond to an equilibrium mutual inclination $78^\circ < \mi < 102^\circ$, which is also unstable in the absence of tides.
We thus conclude that tidal effects only need to be taken into account for close-in planets ($a < 1.5$~au).
In particular, they do not disturb the orbits at the edge of stability ($a>2$~au).
Therefore, the stability islands observed at high eccentricities and inclinations remain a possibility to find planets in close-in binaries.

%%%%%%%%%%%
% a = 0.55; e = 0.907
% a = 1.50; e = 0.960
% a = 2.00; e = 0.968
% a = 3.00; e = 0.977
% a = 4.00; e = 0.982
%%%%%%%%%%%
%k0(0.55,0.907) = 1.10915819628648e-13
%k0(1.5,0.96)   = 5.25797449694907e-11
%
%k0(3,0.977)    = 3.66369477145822e-09
%k0(4,0.982)    = 2.46817573786442e-08
%%%%%%%%%%%%%

\begin{figure*}
\centering
\includegraphics[width=6.0cm]{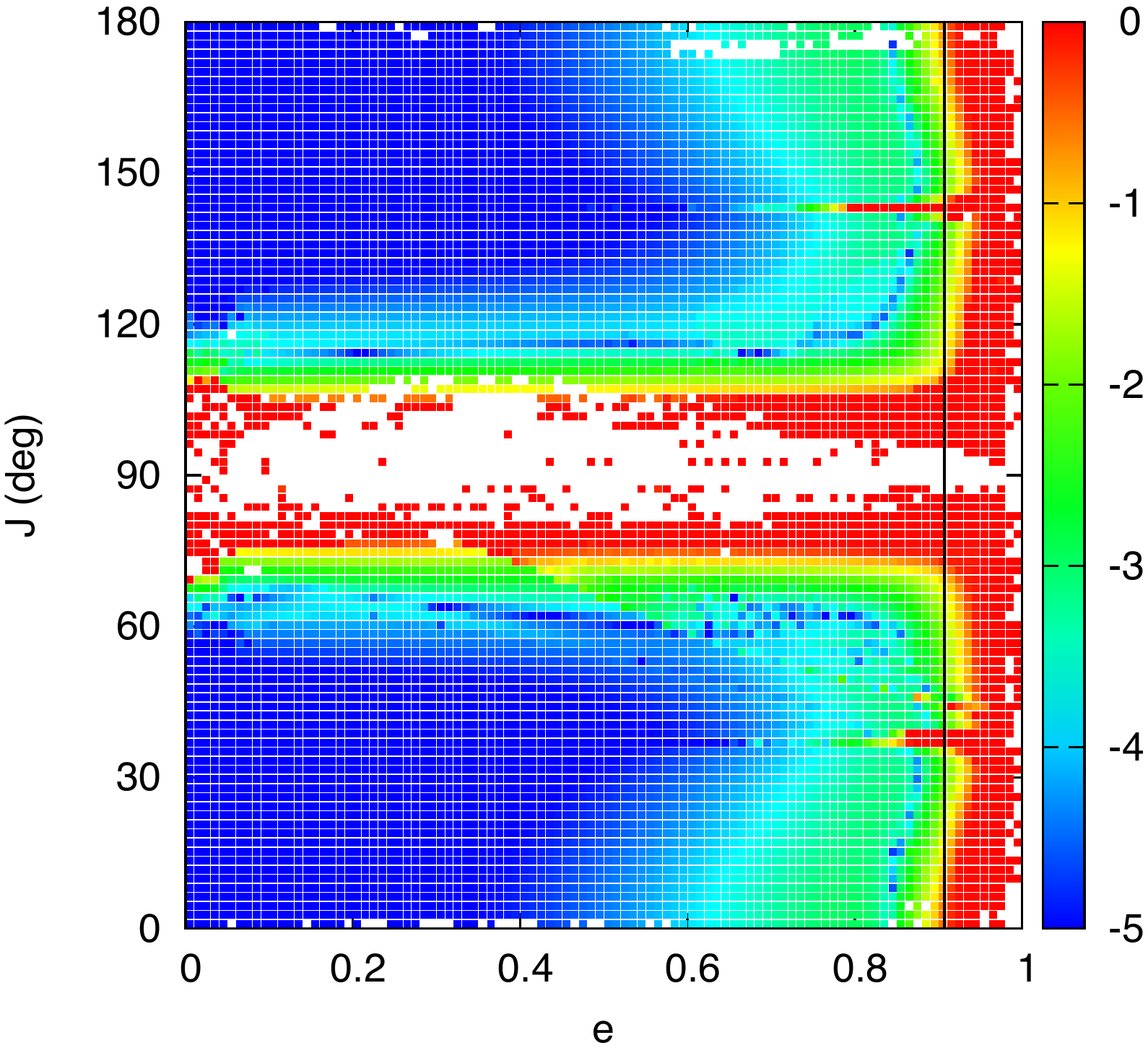}  
\includegraphics[width=6.0cm]{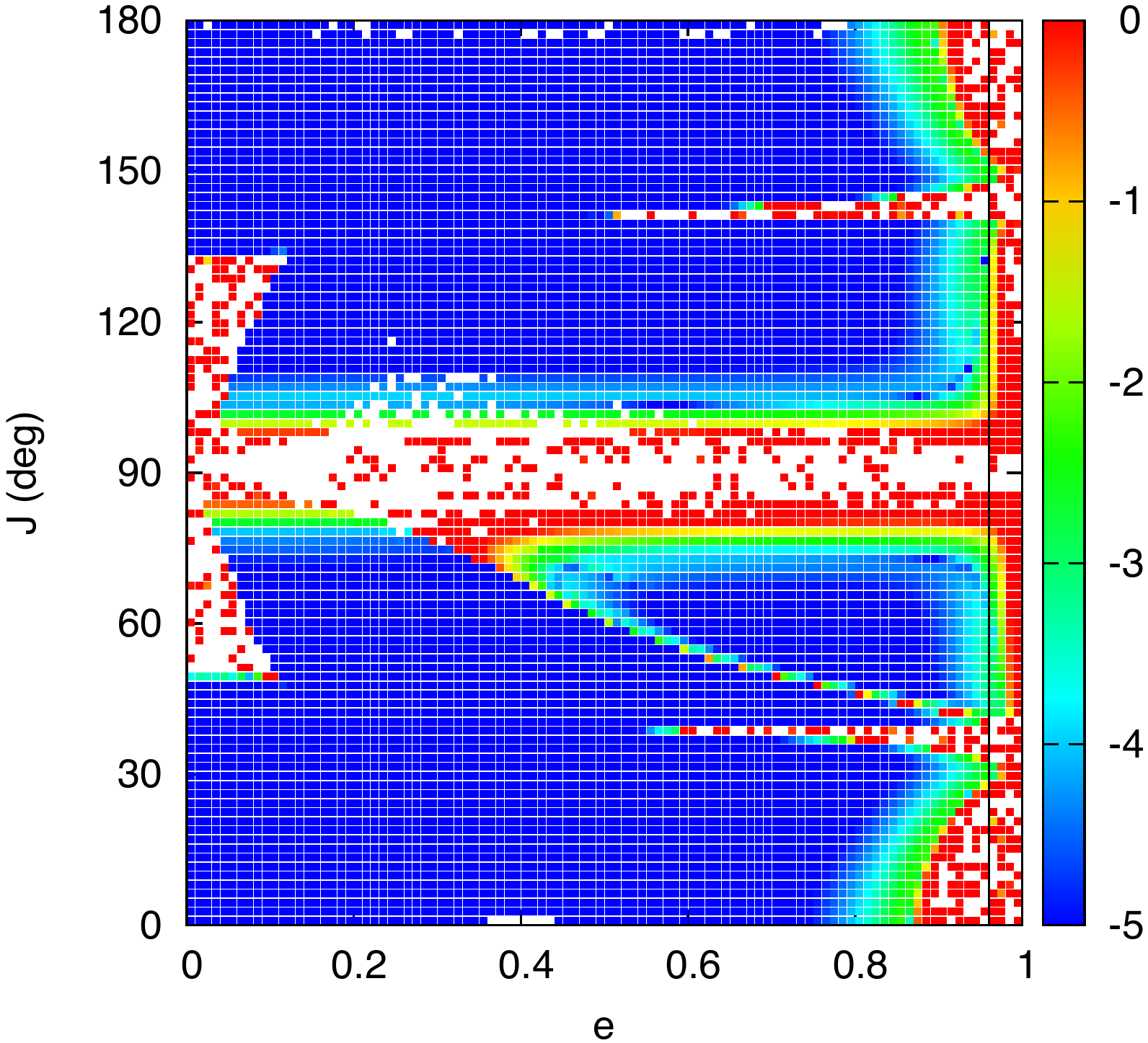} 
\includegraphics[width=6.0cm]{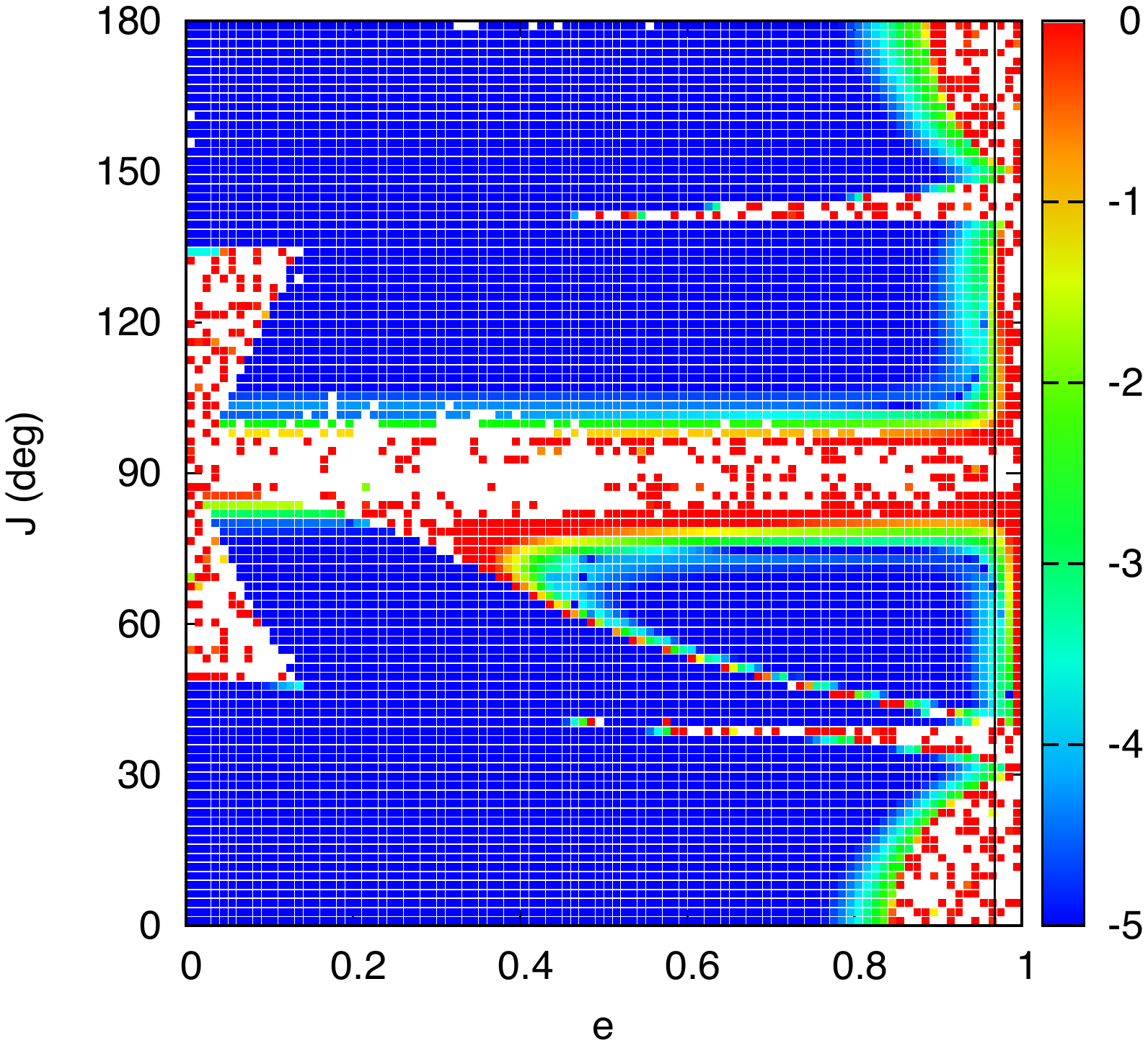}
\caption{Tidal evolution in the plane ($e, \mi$) obtained with a secular model for the same initial conditions from Fig.\,\ref{fig-MapEI}. The colour index gives the variation between the initial and final semimajor axes, $\log_{10} (\Delta a/a)$, and white stands for unstable orbits. The initial semimajor axes from left to right are $a=0.55$, 1.5, and 2.0~au. The black vertical lines give the tidal stability limit (Eq.\,(\ref{100210d})).}
\label{tides:evol}
\end{figure*}

The above equations are only valid for unperturbed orbits, but the initial eccentricity can be seen as the maximal eccentricity over a cycle, so $\tau$ provides a minimal estimation of the dampening time.
Indeed, for orbits inside the Lidov-Kozai resonant region, the orbital damping drives the planet into the exact resonance \citep{Giuppone_etal_2012}, so the eccentricity can stay very high.
A complete analysis requires a study that combines tidal effects with orbital forcing.
{Adopting the secular tidal model\footnote{This model uses the octupolar non-restricted approximation for the orbital interactions, general relativity corrections, the quadrupolar approximation for the spins and the viscous linear model for tides. Although in \citet{correia_etal_2016} the authors apply their model to study P-type circumbinary orbits, it is also valid to study S-type orbits as in \citep{correia_etal_2011}.} from \citet{correia_etal_2016}, we have run simulations for three different semimajor axes with tides using the same initial conditions from Fig.\,\ref{fig-MapEI}.
In Figure~\ref{tides:evol} we show the final evolution of the semimajor axis after 5~Gyr.
The results show that the theoretical estimation given by expression (\ref{100210d}) is relatively accurate and can be used to put constraints on the tidal evolution.
In addition, Fig.\,\ref{tides:evol} also shows that the main chaotic structures are captured by the secular octupolar model.}

\section{Other compact binary systems}
\label{sec.fictitious}
\begin{figure*}
 \centering
\includegraphics[height=4.0cm]{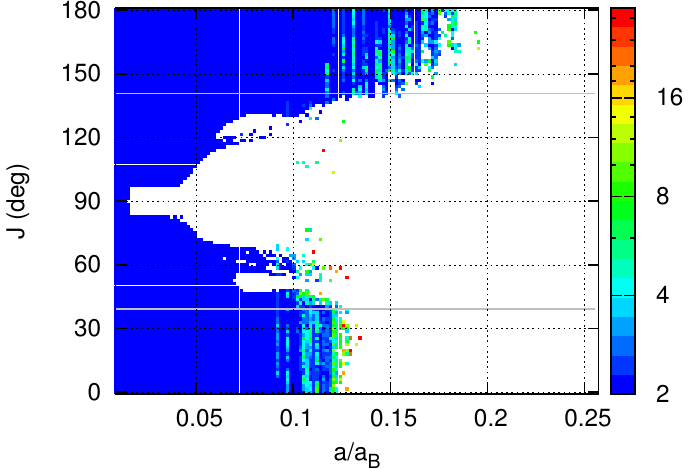}  
\includegraphics[height=4.0cm]{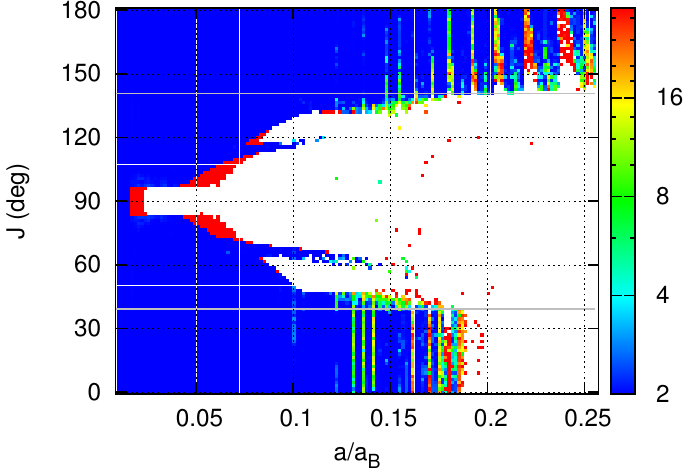} 
\includegraphics[height=4.0cm]{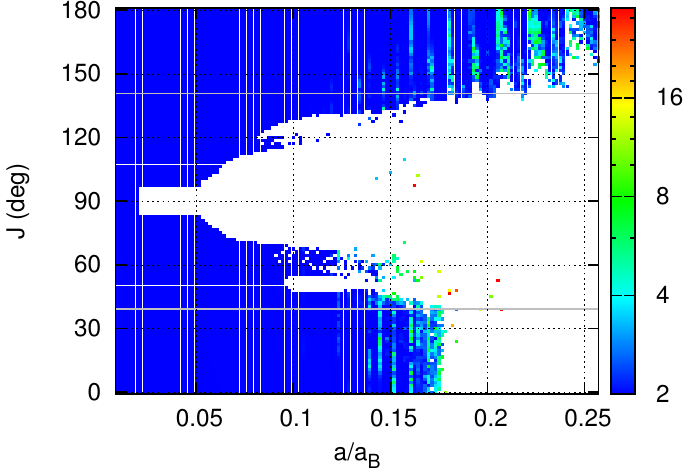}\\ 
\includegraphics[height=4.0cm]{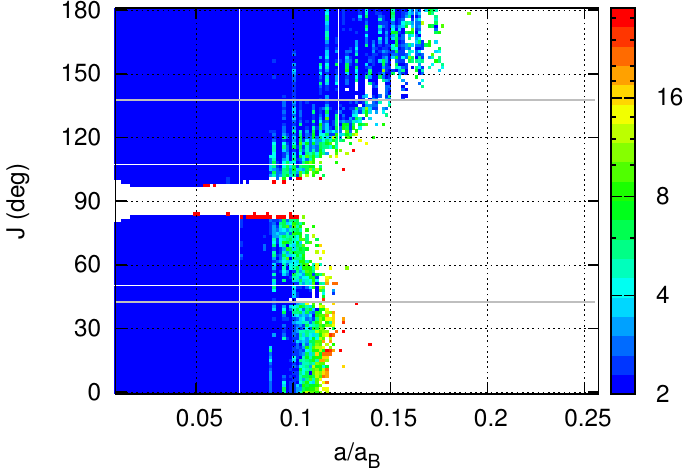}  
\includegraphics[height=4.0cm]{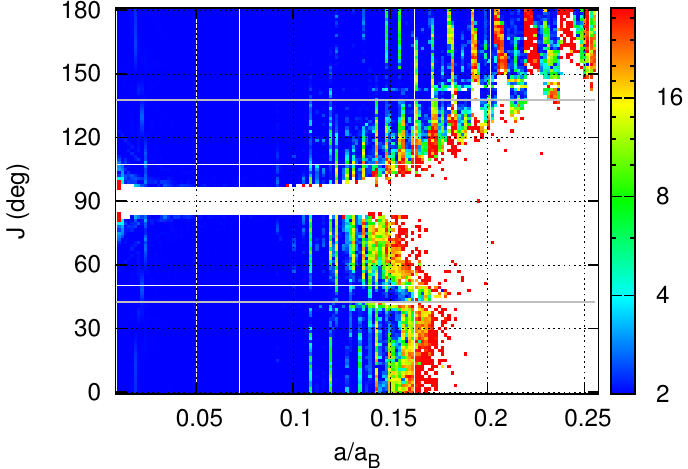} 
\includegraphics[height=4.0cm]{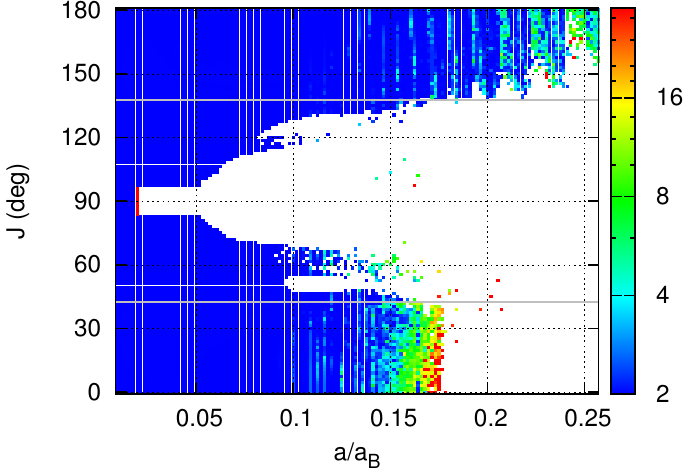}\\
\includegraphics[height=4.0cm]{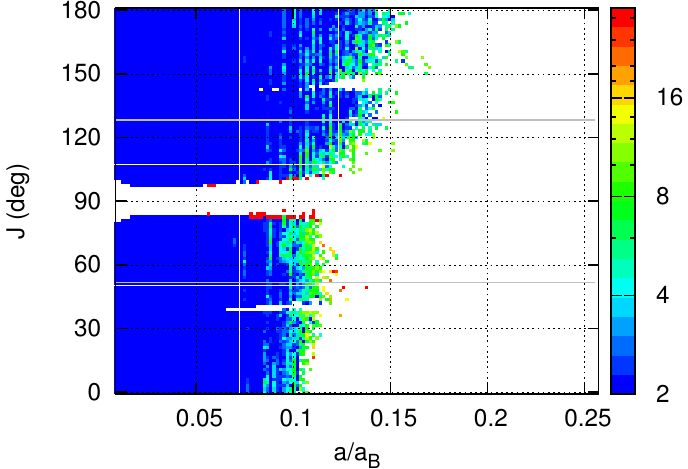}  
\includegraphics[height=4.0cm]{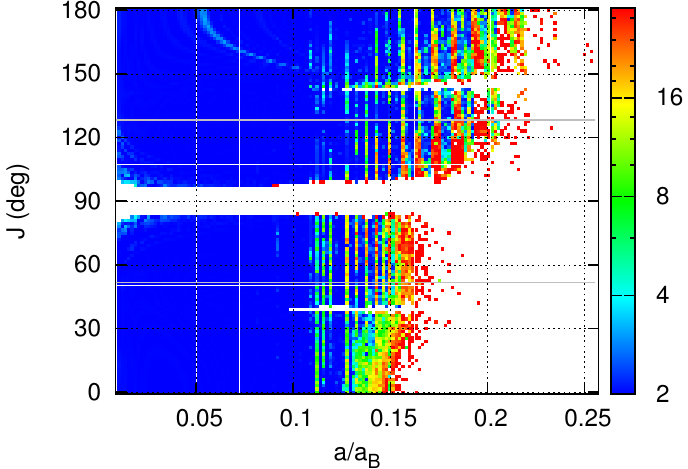} 
\includegraphics[height=4.0cm]{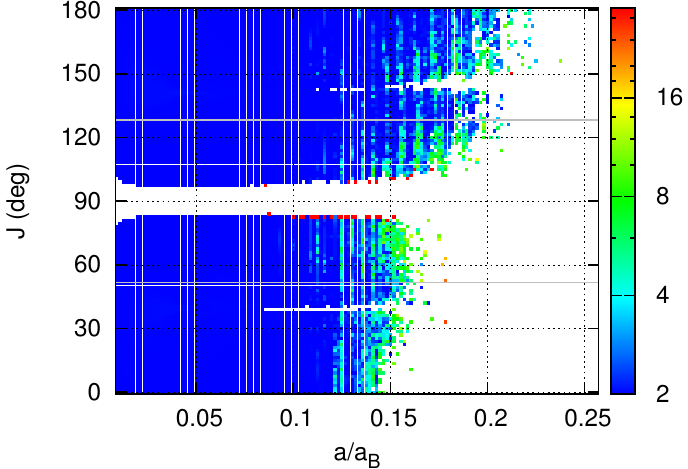}\\
\includegraphics[height=4.0cm]{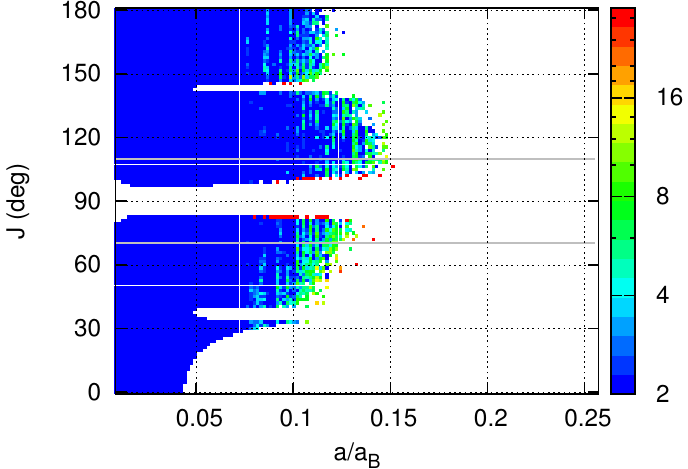}  
\includegraphics[height=4.0cm]{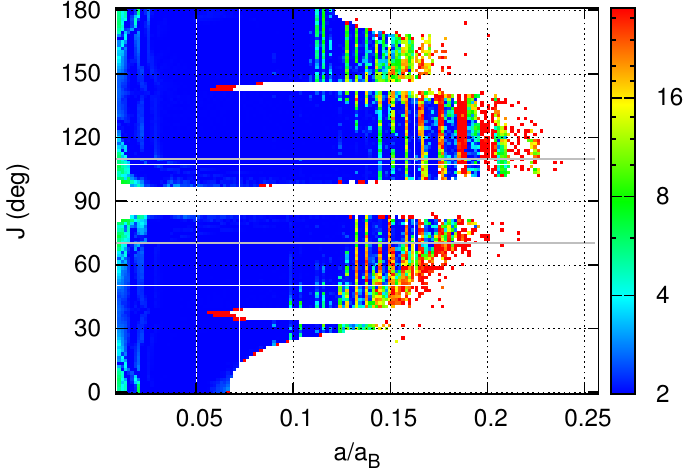}
\includegraphics[height=4.0cm]{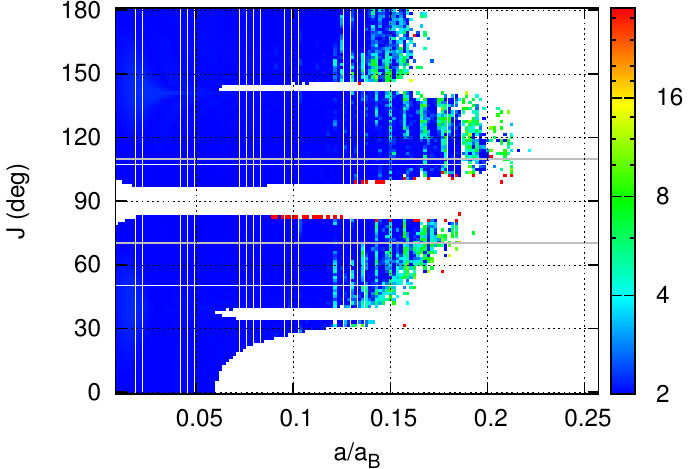}\\
 \caption{Stability maps of \textbf{$\langle Y \rangle$} in the plane ($a/a_B, \mi$) for some eccentricity values, with $\lambda=\Omega=0$ and $\omega=90^\circ$ for three different compact binary systems. From left to right: $\alpha$\,Centauri, HD\,196885, and HD\,41004 are shown. From top to bottom, the initial eccentricity is $e=0$, $0.3$, $0.6$, and $0.9$.}
 \label{fig-ind}
\end{figure*}

We have seen that the Lidov-Kozai resonance is an important mechanism that allows stable regions with high eccentricity and high inclination in the $\alpha$\,Cen system.
Although no planets are known for this system \citep{Hatzes_2013, Endl_2015, Rajpaul_etal_2016}, other similar compact binary systems exist for which planets have been reported in very eccentric orbits such as HD\,196885\,b and HD\,41004\,Ab (see Table~\ref{tab-bin}).
We may then wonder how the stability regions are modified for the different mass ratios of these systems.
As for $\alpha$\,Cen, the stability in the two other binary systems has already been studied before \citep[e.g.][]{Funk_etal_2015}, but they focus on prograde and nearly coplanar systems ($\mi<60 ^\circ$).

In Figure~\ref{fig-ind} we show the stability maps in the plane ($a, \mi$) with $\omega=90^\circ$ for the three compact binary systems listed in Table~\ref{tab-bin}, which include $\alpha$\,Cen. 
These maps are the same as shown in figure~\ref{fig-LK}, but the semimajor axis of the planet is normalised by the semimajor axis of the binary to enable a better comparison between the different systems.
Actually, the semimajor axis of the binary does not change much between the systems, but the mass ratios are $M_B/M_A=0.85$, $0.33,$ and $0.57$ for $\alpha$\,Cen, HD\,196885, and HD\,41004, respectively.
We observe that the results are qualitatively the same with the Lidov-Kozai regions located at same places. %, but shrinking the region of stability.
The only difference is that stability can be obtained for more distant semimajor axis ratios, since the Hill sphere of the main star is larger.
We hence conclude that the stability maps drawn for $\alpha$\,Cen are very general and can be used as reference for other compact binary systems.

\section{Conclusions}
\label{sec.conclusions}

In this paper we have numerically investigated the stability of S-type planetary orbits in the $\alpha$\,Centauri system.
In particular, we studied the stability on inclined orbits for high eccentricities and various orientation angles.

The Hill radius of $\alpha$\,Cen\,A is $\sim 6.4$~au, but stability for coplanar prograde orbits can only be achieved for $a< 3$~au owing to mean-motion resonances overlap (Fig.\,\ref{a:lambda}).
We have shown that nearly coplanar retrograde orbits ($\mi > 150^\circ$) with moderate eccentricity ($e \sim 0.3$) can extend the stability regions beyond 6~au, very close to the limits of the Hill sphere (Fig.\,\ref{fig-LK}).

We have also shown that an exhaustive study of the stability regions cannot be restricted to the action variables ($a, e, \mi$). 
The conjugated angles are also important, in particular the argument of the pericentre $\omega = \varpi - \Omega$.
For simplicity, previous studies usually set $\omega = 0^\circ$, but this choice limits the stability regions at very high inclinations.
For $39.2^\circ < \mi < 140.8^\circ$ large stable regions appear located \tpb{around} $\omega = 90^\circ$ and $\omega = 270^\circ$, corresponding to libration in the Lidov-Kozai resonance.

As the eccentricity increases, the mutual inclination at the centre of the Lidov-Kozai resonance also increases.
The Lidov-Kozai resonant region is thus the most stable region for planets in eccentric orbits.
It persists for high inclinations, but also for semimajor axes close to 4~au.
As for coplanar orbits, the retrograde regions of this resonance $100^\circ < \mi < 140.8^\circ$ are also more stable than the prograde regions $39.2^\circ  < \mi < 80^\circ$.
For very eccentric orbits ($e>0.9$), tidal effects can also modify the Lidov-Kozai equilibrium, but only for  close-in planets ($a<1.5$~au).

%A more global view of the stability regions of S-type planetary orbits is then obtained by fixing the argument of the pericenter at $\omega = 90^\circ$ (or $270^\circ$).

{In this paper we focused on the stability of S-type orbits in the $\alpha$\,Centauri system. 
Nevertheless, our results remain qualitatively the same for any compact binary system with significant eccentricity ($e_B>0.4$).
For binaries in nearly circular orbits, low order resonance capture is possible and the global picture may be different \citep{Marzari_2016}.} 

{Finally, we may wonder about the reliability of forming planets in very eccentric and inclined orbits in binary systems.
Indeed, at present no planets have been found in such configurations. 
However, it appears to be possible to trap single planets at $\sim 2$~au in Lidov-Kozai configurations in compact binary systems ($a_B \sim 20$~au) when tides are considered, through a close fly-by of a background star \citep{marti_beauge_2012, Marti_2015}.}

%%%%%%%%%%%%%%%%%%%%%%%%%%
\begin{acknowledgements}
The authors acknowledge financial support from CIDMA strategic project UID/MAT/04106/2013. 
The computations were performed at the BlaFis cluster at the University of Aveiro.
\end{acknowledgements}

%\appendix
\bibliographystyle{aa}
\bibliography{library}

\end{document}